\documentclass[aps,prb,twocolumn,superscriptaddress,showpacs,floatfix]{revtex4}

\usepackage{amsmath,amssymb}

\usepackage{graphicx}
\graphicspath{{Figures/}}
\DeclareGraphicsExtensions{.eps,.ps}

\begin{document}

\title{
Surface Fluctuations and the Stability of Metal Nanowires
}
\author{C.--H. Zhang}
\affiliation{Department of Physics, University of Arizona, 
1118 E.\ 4th Street, Tucson, AZ 85721}
\author{F. Kassubek}
\affiliation{ABB Schweiz AG, Corporate Research, CH-5405, Baden-D\"attwil, 
Switzerland}
\author{C. A. Stafford}
\affiliation{Department of Physics, University of Arizona, 
1118 E.\ 4th Street, Tucson, AZ 85721}
\

\date{\today}

\begin{abstract}

The surface dynamics and thermodynamics of 
metal nanowires are investigated in a continuum model.
Competition between surface tension and
electron-shell effects 
leads to a rich stability diagram, 
with fingers of stability 
extending to extremely high temperatures 
for certain magic conductance values.
The linearized dynamics of the nanowire's surface are investigated, 
including both acoustic surface phonons and surface self-diffusion of atoms.
On the stability boundary, the surface exhibits critical fluctuations,
and the nanowire becomes inhomogeneous.
Some stability fingers coalesce at higher temperatures, or exhibit overhangs,
leading to reentrant behavior.
The nonlinear surface dynamics of unstable nanowires are also investigated in
a single-mode approximation.
We find evidence that some unstable nanowires do not break, 
but rather neck down to the next stable radius.

\end{abstract}


\maketitle


\section{Introduction}
\label{sec:intro}

Metal wires play an essential role in all electrical circuits, from
power distribution between cities to interconnects in integrated circuits.
In today's technology,
feature sizes down to approximately 100nm are the state
of the art, but current
trends,\cite{ITRS01} consistent with {\em Moore's law},\cite{Moore65}
extrapolate to 1nm technology by 2020.
A question of fundamental importance 
is whether metal
will retain its role as the conductor of choice even at the
ultimate limit of atomic-scale technology, or whether it must be replaced
with more exotic conductors, such as carbon nanotubes.\cite{Dekker99}

A macroscopic analysis of the mechanical properties of thin metal wires
suggests
that it might be difficult to fabricate wires thinner than a few thousand
atoms in cross section:  Consider a cylindrical wire of radius $R$ and
length $L$.  The
maximum tension that the wire can sustain before the onset of plastic flow is
$F_Y = \pi R^2 \sigma_Y$, where $\sigma_Y$ is the {\em yield strength}.
On the other hand, the force due to the surface tension $\sigma_s$ in a thin 
wire is $F_s = - \pi R \sigma_s$.
If $|F_s| > F_Y$, one would expect the
wire to undergo plastic flow and, if $L>2\pi R$, to
break up under surface tension, as in
the {\em Rayleigh instability} of a column of fluid.\cite{Chand81}  
This estimate gives a minimum radius for solidity,
$R_{\rm min}= \sigma_s/\sigma_Y$.  The parameters for several simple metals
are given in Table \ref{table:nanoscale}.
Plateau realized as early as 1873 that
this surface-tension driven instability of a cylinder 
is unavoidable if
cohesion is due solely to classical pairwise interactions
between atoms.\cite{Plateau}

\begin{table}[b]
\begin{center}
\begin{tabular}{|cccccccc|} \hline 
Metal & $\sigma_Y$  & $\sigma_s$  & $\sigma_s(\mbox{\scriptsize FEM})$ &
$\gamma_s$ & $\gamma_s(\mbox{\scriptsize FEM})$ & $\sigma_s/\sigma_Y$
& $G_{\rm min}$ \\
 & (MPa) & (N/m) & (N/m) & (pN) & (pN) & (nm) & ($G_0$) \\ \hline
Cu & 210 & 1.5 & 0.83 & 190 & 140 & 7.1 & 2300 \\
Ag & 140 & 1.0 & 0.51 & 154 & 95 & 7.4 & 1900 \\
Au & 100 & 1.3 & 0.51 & 257 & 96 & 13 & 5600 \\
Li & 15 & 0.44 & 0.37 & 99 & 75 & 29 & 26000 \\
Na & 10 & 0.22 & 0.17 & 39 & 41 & 22 & 10000 \\
\hline 
\end{tabular}
\end{center}
\caption{The yield strength $\sigma_Y$,\cite{Metals:reference} surface energy
$\sigma_s$,\cite{Tyson77} and 
curvature energy $\gamma_s$ \cite{Perdew91} of various monovalent metals.
The values \cite{Stafford99,Stafford00} in the 
free-electron model,
$\sigma_s(\mbox{\scriptsize FEM})=\varepsilon_F k_F^2/80\pi$ and
$\gamma_s(\mbox{\scriptsize FEM})=4\varepsilon_F k_F/45\pi^2$,
are shown for comparison.  For a wire of radius
$R < \sigma_s/\sigma_Y$, the stress due to surface tension exceeds $\sigma_Y$,
signalling a breakdown of macroscopic elasticity theory.  The electrical
conductance $G_{\rm min}$ of a ballistic
wire of radius $R_{\rm min}=\sigma_s/\sigma_Y$ is shown in the rightmost
column, in units of the conductance quantum $G_0=2e^2/h$.
Note that
$G/G_0$ is approximately equal to the number of atoms that fit within the
cross section for monovalent metals.
}
\label{table:nanoscale}
\end{table}

A great deal of experimental evidence
has accumulated over the past decade, however, indicating that metal
nanowires considerably thinner than the above estimate 
can be fabricated by a number of different techniques.
\cite{Rubio96,Untiedt97,Kondo97,Auchains,Yanson99,Yanson01,Ugarte02,Diaz03} 
Even wires with lengths significantly exceeding their circumference
were found to be remarkably stable,\cite{Kondo97,Auchains,Ugarte02} 
indicating that some new mechanism must intervene to prevent their breakup.

An important technique which has been used to model the energetics of metal 
nanowires is 
classical molecular dynamics,
\cite{Landman90,Todorov93,Brandbyge98,Tosatti98,Bil98,Wang01} which utilizes
short-ranged interatomic potentials optimized 
to fit the bulk properties of solids.
This technique has had considerable success,
including predicting the formation of metal nanocontacts in STM experiments
\cite{Landman90} and predicting 
novel, noncrystalline order in nanowires.\cite{Tosatti98}  
However, this approach, which neglects quantum-size effects, 
is unable to avoid the Rayleigh instability in long wires.
\cite{Tosatti98,Bil98,Wang01}

A clue to the resolution of this problem was provided by the observation
of electron-shell structure in conductance histograms of 
alkali metal nanocontacts.\cite{Yanson99}
Like the surface tension, 
quantum-size effects arising from the confinement
of the conduction electrons within the cross-section
of the wire become increasingly important
as the wire is scaled down to atomic dimensions.  In fact,
a linear stability analysis \cite{Kassubek01} 
of metal nanowires within the
free-electron model found that the Rayleigh instability can be completely
suppressed 
for certain favorable radii.

\begin{figure}[t]
\resizebox{8.0cm}{!}{
\includegraphics{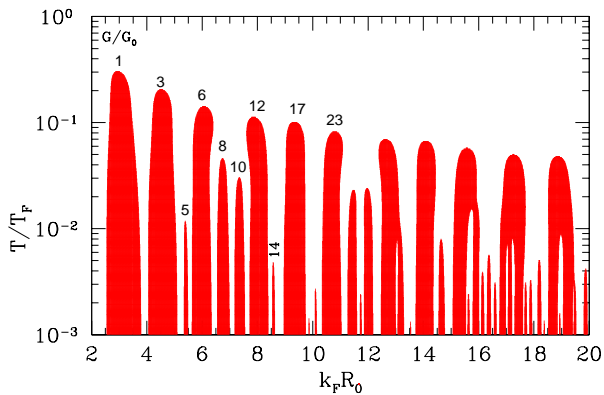}
}
\caption{
Stability of cylindrical metal nanowires as a function of 
radius and temperature.  Shaded regions indicate stability with respect
to small perturbations, $A(R_0,T) > 0$; 
unshaded regions denote unstable configurations, $A(R_0,T)<0$.
Here $T_F$ is the Fermi temperature, $k_F$ the Fermi wavevector, and $R_0$ 
the mean radius of 
the wire.  The quantized conductance values of some of the stable wires 
are indicated.
\label{fig:stability}
}
\end{figure} 

In this article, we investigate the surface dynamics and thermodynamics
of simple metal nanowires in a continuum approach, in 
order to shed further light on their unusual stability properties.
The starting point for our analysis is 
the thermodynamic stability diagram 
shown in Fig.\ \ref{fig:stability}. 
Competition between surface tension and electron-shell effects
leads to a complex landscape of stable fingers and arches extending
up to very high temperatures:  
wires whose electrical conductance is
a magic number 1, 3, 6, 12, 17, 23,...\ times the conductance quantum 
$G_0=2e^2/h$ are predicted to be stable with respect to small perturbations
up to temperatures well above the
bulk melting temperature $T_M \approx .01 T_F$, where $T_F$ is the Fermi
temperature.  
This finding suggests that
metal nanowires may be remarkably robust,
which is cause for optimism about their potential for nanoelectronics
applications.\cite{nanotech} 
Fig.\ \ref{fig:stability} is akin to a {\em phase diagram} for metal 
nanowires;  
the nature of the different phases is revealed in this article
through a study of the surface dynamics for small perturbations about a
cylinder.   We find that the stable fingers correspond to homogeneous
(i.e., translationally invariant) phases, while the intervening
regions correspond to inhomogeneous phases.


\section{The Model}
\label{sec:model}

The continuum model we employ
allows for an analytical treatment of the long-wavelength surface modes 
used to characterize the different phases in Fig.\ \ref{fig:stability},
as well as a correct treatment of
quantum-size effects, which are essential to stabilize long nanowires. 
The ionic degrees of freedom of the wire are modeled as an incompressible,
irrotational fluid, and the conduction electrons are treated as a 
Fermi gas confined within the wire by Dirichlet boundary
conditions at the surface.  
Electron-electron interactions are included only at a macroscopic level (by
requiring the wire to be electrically neutral), since it is well known 
\cite{Brack97,Stafford99,Stafford00}  
that the leading mesoscopic shell-correction to the energy is
independent of interactions.
Calculations including 
interactions at the mean-field level \cite{Yannouleas98b,Puska01} 
yield shell effects very similar to those in the 
free-electron model.\cite{Stafford97,Stafford99}

Modelling the ionic degrees of freedom as a fluid is motivated by the 
argument presented in Table \ref{table:nanoscale}, which indicates that
metal nanowires thinner than a few thousand atoms in cross section 
should be very plastic.  The free-electron model for the conduction electrons
is appropriate to describe electron-shell effects in monovalent metals,
and as the values in Table \ref{table:nanoscale} indicate, even describes
some macroscopic properties of alkali metals semi-quantitatively.
Although the continuum approximation is not justified {\it a priori} in
the limit of atomically-thin wires, 
this model is nonetheless justified {\it a posteriori} by its success
in describing simple metal clusters\cite{clusters} of comparable
dimensions.
Cohesion and quantum transport in gold nanocontacts were also successfully
described with this model.\cite{Stafford97,Buerki99a,Buerki99b}
The directionality of bonding due to contributions from $p$-,
$d$- and $f$-electrons is 
of course absent from the free-electron model, as are element-specific 
effects, such as the tendency toward surface reconstruction, which was 
argued to play an essential role in the formation of atomic 
chains.\cite{Smit01}  Nonetheless, wires with $G=G_0$ are predicted to
be very stable within the free-electron model, 
and the strength of a metallic bond in such a wire
is significantly greater than that in the bulk.\cite{Stafford97,Stafford99}

An empirical justification for our continuum model 
comes from experimental results indicating that 
electron-shell effects dominate over ionic ordering in sufficiently
thin alkali metal\cite{Yanson01} and gold\cite{Diaz03} wires.  
Yanson {\it et al.}\cite{Yanson01} found an interesting interplay between
electron-shell effects and atomic-shell effects in alkali metal nanocontacts.
Electron-shell effects were found to be most important in the lighter
elements lithium and sodium, presumably due to the larger Fermi energies of the 
conduction electrons and the lighter, more mobile ions, while atomic-shell
effects were most important in the heavier element potassium.  A crossover
from electron-shell structure to atomic-shell structure in conductance
histograms was found for conductance values $G/G_0\approx 36$ in 
potassium, while electron-shell effects were found to dominate even for
$G/G_0 > 100$ in lithium.  An intermediate behavior was observed for sodium.
Interestingly, the competition between the two effects was found to be 
history dependent.  In a particular sequence of histograms obtained by
cycling a potassium break junction, an evolution from atomic-shell structure
to electron-shell structure was observed.\cite{Yanson01}  Most recently,
a similar interplay between electron-shell structure and atomic-shell
structure was also observed in gold nanocontacts.\cite{Diaz03}

These fascinating experimental
results cry out for deeper theoretical investigations of 
the stability and structure of metal nanowires.
While the geometry of the nanowires studied in Refs.\ 
\onlinecite{Yanson99,Yanson01,Diaz03} was not directly determined, they
may be rather short due to the fabrication method, 
so that the connection\cite{Buerki02} 
to the contacts may play an important role.  
In this article, 
we study the more theoretically tractable---and more technologically 
relevant---problem of the stability and surface dynamics of long metal
nanowires.  Our analysis should be directly relevant for the nanowires 
studied in Refs.\ \onlinecite{Kondo97,Ugarte02}.

\section{Linear stability analysis}
\label{sec:stability}

In our continuum 
model, the ionic degrees of freedom are completely determined by
the surface coordinates of the wire.  
Motivated by the fact that
only modes which preserve axial symmetry participate in the surface-tension
driven instability of a cylinder,\cite{Chand81} we restrict our 
consideration to axially-symmetric perturbations, 
\begin{equation}  
R(z,t)=R_0+\sum_n b(q_n,t)e^{iq_n z},
\label{eq:pert}
\end{equation}
where $R(z,t)$ is the radius of the wire at position $z$ and time $t$, 
$R_0$ is the unperturbed radius,
and $b(q,t)=b^*(-q,t)$ are complex Fourier coefficients. 
Periodic boundary conditions are assumed for a wire of length $L$, so that
$q_n=2\pi n/L$, with $n$ an integer bounded by
$|n| \leq N \approx k_F L/\pi$ (a lattice cutoff).
Since the total number of atoms comprising the nanowire is unchanged by
the perturbation, $b(0,t)$ is
related to the other $b(q_n,t)$ by volume 
conservation 
\begin{equation}
b(0,t)+\frac{b^2(0,t)}{2R_0}=-\frac{1}{R_0}\sum_{n=1}^N |b(q_n,t)|^2,
\label{eq:b0}
\end{equation}
and may be eliminated.  

\begin{figure}
\begin{center}
\resizebox{6.0cm}{!}{
\includegraphics{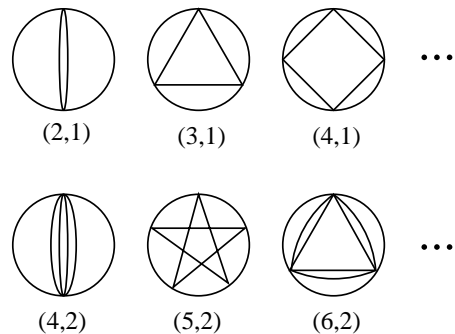}
}
\end{center}
\caption{Classical periodic orbits \cite{Balian72,Brack97}
of an electron in a plane perpendicular to the axis of the wire,
labeled $(v,w)$,
where $v$ is the number of vertices and $w$ is the winding number.
}
\label{fig:orbits}
\end{figure}

For small perturbations, the grand canonical potential of the electron
gas is quadratic in the Fourier coefficients $b(q,t)$, and determines  
the potential energy $U$ 
of the ions in the Born-Oppenheimer approximation,
\begin{equation}
U 
=  U_0(R_0,T)+ L \sum_{n =1}^N
 \alpha(q_n;R_0,T) |b(q_n,t)|^2,
\label{eq:potential}
\end{equation}
where $U_0(R_0,T)$ is the potential energy
of an unperturbed cylinder,
\begin{equation}
\frac{U_0(R,T)}{L}= \pi R^2 u + 2 \pi R \sigma_s - \pi \gamma_s + V(R,T).
\label{eq:U0}
\end{equation}
Here $u$ 
is the macroscopic free energy density of the electron gas, 
$\sigma_s$ is the surface tension, 
$\gamma_s$ is the surface curvature energy 
(c.f.\ Table \ref{table:nanoscale}), and $V$ is a mesoscopic electron-shell
correction. 
The mode stiffness $\alpha(q;R_0,T)$ has the following form 
\cite{Kassubek01} 
in the semiclassical approximation,
valid for long-wavelength perturbations:
\begin{eqnarray}
\alpha(q;R,T) & = & - 2\pi \sigma_s/R +
2\pi(\sigma_s R - \gamma_s)q^2 
\nonumber \\
& & +
\left(\frac{\partial^2}{\partial R^2}
-\frac{1}{R}\frac{\partial}{\partial R}\right)V(R,T),
\label{eq:alpha}
\end{eqnarray}
where
\begin{equation}
V(R,T) 
= \frac{2\varepsilon_F}{\pi}
\sum_{w=1}^\infty \sum_{v=2w}^\infty
\frac{a_{vw}(T) f_{vw}}{v^2L_{vw}}
\cos(k_F L_{vw}-3v\pi/2).
\label{eq:gutzwiller}
\end{equation}
The sum in Eq.\ (\ref{eq:gutzwiller})
includes all classical periodic orbits $(v,w)$
in a disk billiard (see Fig.\ \ref{fig:orbits}).
$L_{vw}=2vR\sin(\pi w/v)$ is the length of an orbit,
the factor $f_{vw}=1$ for $v=2w$, 2 otherwise,
accounts for the invariance under
time-reversal symmetry of some orbits, and
$a_{vw}(T) = \tau_{vw}/\sinh{\tau_{vw}}$
($\tau_{vw}=\pi k_FL_{vw}T/2T_F$) is a temperature-dependent
damping factor.\cite{Balian72,Brack97}

Since $\partial^2 \alpha/\partial q^2 >0$
for all physically meaningful radii,
long-wavelength perturbations cost the least energy,\cite{Urban}
and the stability of the wire is determined by the sign of 
$A(R_0,T) \equiv \alpha(q=0;R_0,T)$.
For $A(R_0,T)>0$, a nanowire is stable with respect to all small 
perturbations, and is hence a metastable thermodynamic state.  
For $A(R_0,T) < 0$, the nanowire is unstable.  
The stability diagram
so determined is shown in Fig.\ \ref{fig:stability}.
In Fig.\ \ref{fig:stability}, the values $\sigma_s=\sigma_s(\rm FEM)$
and $\gamma_s = \gamma_s(\rm FEM)$, appropriate for alkali metals, were used 
(c.f.\ Table \ref{table:nanoscale}).  For larger values of $\sigma_s$
(e.g.\ for noble metals), the maximum temperatures
(in units of $T_F$) of the stable fingers
are reduced somewhat, but the stability diagram is qualitatively similar.

Further insight into the stability criterion $A>0$ is provided by
the identity
\begin{equation}
A(R_0,T) = \left(\frac{\partial^2}{\partial R_0^2}
-\frac{1}{R_0}\frac{\partial}{\partial R_0}\right)\frac{U_0(R_0,T)}{L}.
\end{equation}
The wire can lower its potential energy via
a volume-conserving
separation into thicker and thinner segments
if and only if $A<0$.
$A<0$ thus corresponds to an {\em inhomogeneous} phase, while $A>0$ 
corresponds to a {\em homogeneous} phase.

Our analysis of stability in terms of the convexity
of the constrained energy functional is quite different from simply
comparing the energy of cylinders of different radius,
\cite{Tosatti98,Puska01,Tosatti01}
which does not address the fundamental question: {\em whether 
any cylinder is stable}.
We also point out that for a sufficiently large system, the 
number of atoms is conserved---neglecting sublimation---and 
the depletion of atoms from
a finite segment of wire \cite{Tosatti01} can be 
described as a finite-wavelength perturbation of a larger system.

Note that our stability analysis is carried out at fixed $L$.  
The tensile force necessary to fix the length of the wire is given by
$F=-\partial U_0/\partial L$ [plus a small correction due to surface
fluctuations, c.f.\ Eq.\ (\ref{eq:free_energy})], and was previously calculated as a function of
radius in this model in Refs.\ \onlinecite{Stafford97,Stafford99,Stafford00}.
Our stability analysis is thus appropriate to describe nanowires under tensile
stress, such as those studied in the experiments of Refs.\
\onlinecite{Rubio96,Untiedt97,Kondo97,Auchains,Yanson99,Yanson01,Ugarte02,Diaz03}.
The stability of a nanowire with free ends is an open question.

\section{Linearized surface dynamics}
\label{sec:dynamics}

\subsection{Surface phonons}
\label{sec:phonons}

We first consider inertial dynamics of the ionic background.
Assuming that the ionic medium is irrotational and incompressible, 
\cite{Yannouleas98a} 
its velocity distribution $\vec{v}(\vec{r},t)$
can be written in terms of a potential satisfying 
the Laplace equation
\begin{equation}
\nabla^2\Phi(\vec{r},t)=0,
\end{equation}
where $\vec{v}(\vec{r},t)= - \nabla \Phi(\vec{r},t)$.
The general solution to this equation with axial symmetry, which is regular 
at $r=0$, can be written\cite{Chand81}
\begin{equation}
\Phi(\vec{r},t)=\Phi(r,z,t)=\sum_{n=-N}^N d(q_n,t)I_0(q_nr)
e^{iq_nz},
\label{eq:velocity_potential}
\end{equation}
where $I_0$ is the modified Bessel function of order zero and 
$r$ is the distance of an 
ion from the $z$-axis. 

For small deformations, the relation between the coefficients $d(q_n,t)$ in
the expansion (\ref{eq:velocity_potential}) 
and the Fourier coefficients $b(q_n,t)$ of the surface perturbation 
(\ref{eq:pert})
can be determined by the condition that the radial 
component of the velocity at the surface is
\begin{equation}
v_r=-\frac{\partial\Phi(r,z,t)}{\partial r}|_{r=R_0}
=\frac{\partial R(z,t)}{\partial t}
\end{equation}
plus terms ${\cal O}(b^3)$.  Therefore, we have 
\begin{equation}
d(q_n,t)=-\frac{1}{q_n I_1(q_nR_0)}\frac{\partial b(q_n,t)}{\partial t},
\label{eq:dandb}
\end{equation}
where $I_1$ is the first-order modified Bessel function. 
The kinetic energy of the ionic medium is then given by
\begin{eqnarray}
K & = &
\frac{\rho_i}{2}\int d^3 r \, \nabla\Phi^\ast(\vec r,t)\cdot 
\nabla\Phi(\vec r,t) 
\nonumber \\
& = & L \sum_{n=1}^N  m(q_n,R_0)\left|\frac{\partial b(q_n,t)}{
\partial t}\right|^2,
\label{eq:ke}
\end{eqnarray}
where $\rho_i$ is the ionic mass density, and 
\begin{equation}
m(q,R)= \rho_i \frac{2\pi R \, I_0(qR)}{qI_1(qR)}.
\label{eq:mass}
\end{equation}
Details of the derivation of Eqs.\ (\ref{eq:ke}) and (\ref{eq:mass}) are
given in Appendix \ref{sec:appendix}.
Combining Eqs.\ (\ref{eq:potential}) and (\ref{eq:ke}) yields a Hamiltonian
for surface phonons, with frequencies
\begin{equation}
\omega(q;R_0,T)=\sqrt{\frac{\alpha(q;R_0,T)}{m(q,R_0)}}.
\label{eq:omega}
\end{equation}
Generically, $\omega(q) \propto q$ as $q\rightarrow 0$ due to the 
$q$-dependence of $m(q,R_0)$. Eq.\ (\ref{eq:omega}) thus describes
{\em acoustic surface phonons}.  On the stability boundary
$A=0$, one has $\omega(q) \propto q^2$ as $q \rightarrow 0$.
For $A < 0$, $\omega$ is imaginary, and long-wavelength modes 
grow exponentially [see Fig.\ \ref{fig:unstable}(a)]. 

\subsection{Surface diffusion}
\label{sec:diffusion}

The surface deformation (\ref{eq:pert}) also 
produces a gradient in chemical potential that drives 
the surface atoms to diffuse---a process likely to be important for
large-scale deformations.\cite{Buerki02} 
The surface current of atoms is given by Fick's law 
\begin{equation}
\label{flux}
\vec{J}=-\frac{\rho_sD_s}{k_BT}\nabla\mu,
\end{equation} 
where $\rho_s$ is the surface density of atoms and $D_s$ is the surface
self-diffusion constant.
Using the continuity equation for the 
surface current, Eq.~(\ref{flux}) can 
be converted into a (linearized) equation of motion for the profile $R(z,t)$  
\begin{equation}
\frac{\partial R(z,t)}{\partial t}
=\frac{\rho_sD_s v_a}{k_BT}\frac{\partial^2\mu}{\partial z^2},
\label{eq:ofmotion_diff}
\end{equation}
where $v_a = 3\pi^2/k_F^3$ is the volume of an atom.
The chemical potential $\mu$ of an atom
is obtained by calculating the change in free energy 
with the addition of an atom at point $z_0$,
\begin{equation}
\mu(z_0,t)  =  U[R(z,t)+C\delta(z-z_0)] - U[R(z,t)],
\label{eq:mu_def}
\end{equation}
where $C=v_a/2\pi R$ is chosen so that the volume of an atom is added.
From Eq.\ (\ref{eq:potential}), one obtains
\begin{equation}
\mu(z,t) = \mu_0+\frac{\varepsilon_F \, v_a}{\pi R_0}
\sum_{n=-N}^N \alpha(q_n;R_0,T)) b(q_n,t)e^{iq_n z},
\label{eq:mu_linear}
\end{equation}
where $\mu_0(R_0,T)$ is the chemical potential of the unperturbed cylinder.
Combining Eqs.\ (\ref{eq:ofmotion_diff}) and (\ref{eq:mu_linear}) yields an
equation of motion for the Fourier component $b(q,t)$,
\begin{equation}
\label{eq:surface_diff}
\frac{\partial b(q,t)}{\partial t}=-\Gamma(q;R_0,T)b(q,t),
\end{equation}
where the relaxation rate
\begin{equation}
\Gamma(q;R_0,T) = \frac{\rho_s D_s v_a^2}{\pi R_0 k_B T}\, q^2 \alpha(q;R_0,T).
\label{eq:gamma}
\end{equation} 
Thus, one finds that under surface diffusion alone, 
a perturbed metastable wire relaxes exponentially toward a cylindrical shape. 
For $\alpha < 0$, the mode grows exponentially.  

\subsection{Combined dynamics}
\label{sec:combo}

Combining inertial and diffusive processes,
the linearized (classical) equation of motion for the surface modes is
\begin{equation}
\frac{\partial^2 b(q,t)}{\partial t^2}
+\Gamma(q)\frac{\partial b(q,t)}{\partial t}+\omega^2(q)b(q,t)=0. 
\label{eq:ofmotion}
\end{equation} 
From the $q$-dependence of Eqs.\ (\ref{eq:omega}) and (\ref{eq:gamma}),
one sees that $\Gamma(q)/\omega(q) \rightarrow 0$ 
as $q\rightarrow 0$, indicating
that diffusive processes can be neglected in this limit, at least for
small deformations.  
In general, the relative time scales for 
inertial and diffusive dynamics depend on the value of $D_s$.
For this quasi-one-dimensional diffusion problem, one can estimate
$D_s \sim (\omega_D/\rho_s)\exp(-E_s/k_B T)$, where $\omega_D$ is
the Debye frequency and $E_s$ is the activation energy for surface 
diffusion, which is comparable to the energy of a single bond in the 
solid. 
With this estimate, one has $\omega(q) \gg \Gamma(q)$
for all $q$, indicating that the surface phonons are underdamped.  

The picture of the surface dynamics of (meta)stable metal nanowires
which emerges from this analysis is that there is a separation of timescales:
on short timescales, the surface oscillates rapidly about the cylindrical
equilibrium shape, while on much longer timescales, surface atoms diffuse 
irreversibly.


\section{Critical surface fluctuations} 
\label{sec:rough}

In the harmonic approximation [Eqs.\ (\ref{eq:potential}) and
(\ref{eq:ke})], the total free energy $\Omega(R_0,T)$ 
of the nanowire is given by the 
free energy of the unperturbed cylinder plus the Helmholtz free energy of 
the surface phonons,
\begin{equation}
\Omega = U_0 + \sum_{n=1}^N \left[\hbar \omega_n + 2 k_B T 
\ln \left(1-e^{-\beta \hbar \omega_n}\right)\right],
\label{eq:free_energy}
\end{equation}
where $\omega_n \equiv \omega(q_n;R_0,T)$ and $\beta=1/k_B T$.  The equilibrium
tension in the wire is 
\begin{equation}
F=-\frac{\partial \Omega}{\partial L}=-\frac{\partial U_0}{\partial L}
+ \delta F_{\rm phonon}, 
\label{eq:tension}
\end{equation}
where the main contribution $-\partial U_0/\partial L$ was previously 
calculated in Refs.\ \onlinecite{Stafford97,Stafford99,Stafford00}, and
$\delta F_{\rm phonon}$ is a small correction
that is singular at the stability boundaries, where the surface modes
become soft.

\begin{figure}
\resizebox{8.0cm}{!}{
\includegraphics{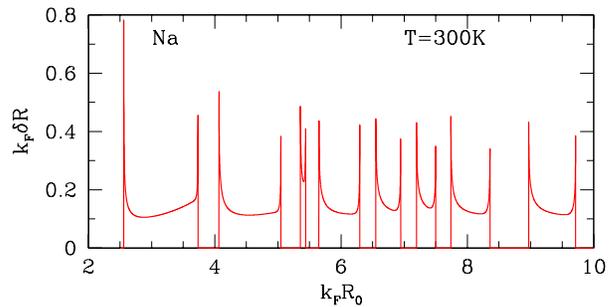}}
\caption{Root-mean-square fluctuations of the radius
of metastable sodium nanowires of length $L=10R_0$. 
$\delta R$ is undefined for unstable wires within the harmonic
approximation, and is not shown.}
\label{fig:rough}
\end{figure}

The softening of the surface modes on the stability boundaries leads to
critical surface fluctuations.
Given the stiffness (\ref{eq:alpha}) and frequency (\ref{eq:omega}) of
the surface modes, 
the mean-square thermal fluctuation $\delta R^2$ of the radius of the nanowire 
can be calculated in the usual way\cite{Fisher83}
\begin{equation}
\delta R^2 \equiv \langle(R-R_0)^2\rangle
=\frac{1}{L}\sum_{n=1}^N \frac{\hbar\omega_n
[2f(\omega_n) + 1]}{\alpha(q_n;R_0,T)},
\label{eq:roughdef}
\end{equation}
where $f(\omega)=[\exp(\beta\hbar\omega)-1]^{-1}$ is the Planck distribution.
Note that, aside from a small quantum correction, the magnitude of the 
surface fluctuations
follows from the equipartition theorem applied to Eq.\ (\ref{eq:potential}), 
and is thus largely independent of the 
nature of the surface dynamics---whether inertial or diffusive.

Fig.\ \ref{fig:rough} shows $\delta R$ for nanowires of finite length 
at room temperature as a function of their mean radius.
The ionic mass and Fermi temperature were taken to be that of sodium.
Within a metastable region, 
$k_F \delta R \ll 1$ and is approximately independent of $L$,
indicating that such 
wires are nearly atomically smooth at this temperature.  
The zero-point motion contributes roughly fifty
percent of the surface fluctuation within a stable region for sodium
nanowires at $T=300K$.  With increasing temperature,
the thermal contribution to $\delta R$ grows proportional to
$\sqrt{T}$, according to the equipartition theorem.  Note that the harmonic
approximation is expected to break down when $k_F \delta R \sim 1$.

The surface fluctuation $\delta R$ exhibits
sharp peaks at the stability boundaries, reaching the value  
\begin{equation}
\left.\delta R^2\right|_{A(R_0,T)=0} = \frac{k_B T}{4 \pi^3 (\sigma R_0
-\gamma)}L,
\label{eq:dRcrit}
\end{equation} 
plus a small quantum correction.
$\delta R$ thus scales as $L^{1/2}$ on the stability 
boundary, like the finite-size scaling at the roughening 
transition of a planar interface.\cite{Fisher86}  
At $T=0$, $\delta R$ remains small and approximately independent of
$L$ on the stability boundary.
The absence of critical surface fluctuations at $T=0$ is also consistent
with the behavior of planar interfaces.\cite{Fisher83}  

The stability boundary $A(R_0,T)=0$ defines a (multiple-valued)
critical temperature $T_c = T_c(R_0)$ as a function of the mean radius, 
or alternatively a critical mean radius $R_c = R_c(T)$ as a function of
temperature (see Fig.\ \ref{fig:stability}).  
Within the harmonic approximation, 
$\delta R$  
grows with an exponent 
$\nu = -1/4$ as $R_0 \rightarrow R_c$ or $T\rightarrow T_c$,
as expected from the Ornstein-Zernicke fluctuation theory.  This critical
behavior, which is cut off when $\delta R$ approaches the value given in
Eq.\ (\ref{eq:dRcrit}),
is illustrated in Fig.\ \ref{fig:rough}. 

One can also study the time-dependence \cite{Flekkoy95} 
of $\delta R$ for an initially cylindrical wire undergoing
thermal fluctuations. Using the classical equipartition theorem, for each 
mode we have 
\begin{equation}
\langle|b(q_n,t)|^2\rangle=\frac{2k_BT}{L}
\frac{\sin^2(\omega_n t)}{\alpha(q_n;R_0,T)}
\end{equation}
where $\sin^2(\omega_n t)$ describes the standing capillary waves. The 
surface fluctuations then grow as a function of time according to
\begin{equation}
\delta R^2(t)=\frac{4k_BT}{L}\sum_{n=1}^N
\frac{\sin^2(\omega_n t)}{\alpha(q_n;R_0,T)}.
\end{equation}
On the stability boundaries $A(R_0,T)=0$, the dispersion relation 
$\omega(q)\sim q^2$, and one finds asymptotically
$\delta R(t) \sim t^{1/4}$  
for times $\omega_1^{-1} \gg t \gg \omega_N^{-1}$. 
The dynamic exponent $z=1/4$ differs from that of a planar interface
\cite{Flekkoy95} due to the different dispersion relation for
the surface modes.

However, all these scaling relations hold only in a limited range,
since the asymptotic limit
$\delta R \rightarrow \infty$ characterizing the roughening
transition\cite{Fisher83,Fisher86,Flekkoy95} 
is unphysical in nanowires, due to their finite radius.
On the stability boundary, 
the surface does not {\em roughen} in a thermodynamic sense, but 
the nanowire does become inhomogeneous.

\section{Reentrant behavior}
\label{sec:reentrant}

\begin{figure}
\resizebox{8.0cm}{!}{
\includegraphics{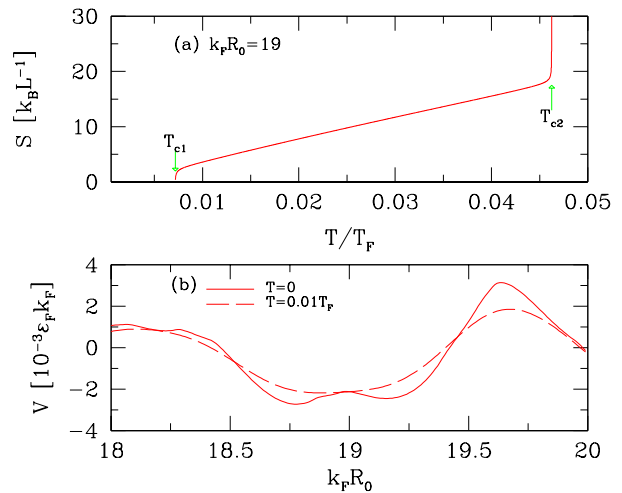}}
\caption{(a) The total entropy per unit length of 
a nanowire with $k_F R_0=19$ versus temperature. 
The ionic mass was taken to be that of sodium.
(b) The electron-shell potential $V(R_0,T)$, from Eq.\ (\ref{eq:gutzwiller}). 
}
\label{fig:entropy}
\end{figure}
 
Perhaps most interesting is the {\em reentrant} behavior
occuring on the arches and overhangs in the 
stability diagram, Fig.\ \ref{fig:stability}.  
For instance, a wire with $k_F R_0 =19$ 
is metastable and homogeneous
in the temperature interval $T_{c1} < T < T_{c2}$,
with $T_{c1} \approx .0072 T_F$ and $T_{c2} \approx .046 T_F$.  
The surface exhibits critical fluctuations
as $T\rightarrow T_{c1}^+$ or
$T\rightarrow T_{c2}^-$, at which points
 the wire makes a transition to an inhomogeneous
phase.  The transition at $T_{c2}$ is
conventional, in the sense that the inhomogeneous phase is the high-entropy
phase.  However, the inhomogeneous phase below $T_{c1}$ has {\em lower entropy}
than the homogeneous phase above $T_{c1}$.  
Fig.\ \ref{fig:entropy}(a) shows 
the total entropy $S=-\partial \Omega/\partial T$
of the nanowire as a function of temperature,
including both electron and phonon contributions, where $\Omega$ is given
by Eq.\ (\ref{eq:free_energy}).  The electronic
entropy is regular at the critical points, but the phonon entropy
is singular in the harmonic approximation, due to the emergence of soft
surface modes. 
The singular contribution to the phonon entropy is
\begin{equation}
S_{\rm sing}(R_0,T) = -\frac{\hbar}{L} \sum_{n=1}^N f(\omega_n) 
\frac{\partial \omega(q_n; R_0, T)}{\partial T},
\label{eq:S_singular}
\end{equation}
indicating that
the softening of a phonon mode with decreasing temperature indeed leads to a 
decrease in entropy.

To understand the counterintuitive behavior at $T_{c1}$, it
is useful to consider the electron-shell correction $V$ to 
the energy of the wire, 
shown in Fig.\ \ref{fig:entropy}(b).   Above
$T_{c1}$, $V$ has a single broad minimum near $k_F R_0 =19$,
but as the temperature is lowered, and the fine structure in the 
shell potential emerges, this single minimum splits into two
minima at $k_F R_0 =18.75$ and $k_F R_0 = 19.2$.  To lower its
free energy, the system would like to fall into one of these two
minima, but due to volume conservation, such a global change is not 
possible.  The wire thus undergoes phase separation into thick and 
thin segments.\cite{Buerki02}

\section{Unstable wires}
\label{sec:unstable}

\begin{figure}[b]
\resizebox{8.0cm}{!}{
\includegraphics{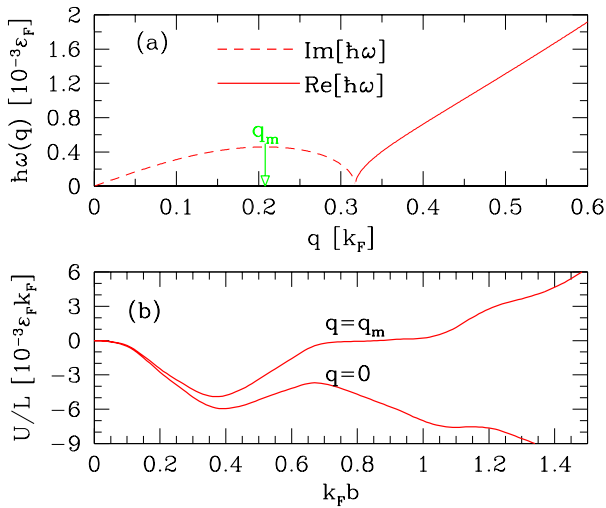}}
\caption{(a) The dispersion relation for the surface modes
of an unstable nanowire with $k_F R_0=8.95$ at $T=0$. 
The ionic mass was taken to be that of sodium.
(b) The potential energy $U[b(q)]$
of the same nanowire for sinusoidal deformations
with $q=0$ and $q=q_m=0.208 k_F$, respectively.
Note that the energy $U_0$ of a straight wire has been subtracted.
The different energies
of these two modes is mainly due to the increased 
surface energy at finite $q$.}
\label{fig:unstable}
\end{figure}

Finally, let us discuss the dynamics of unstable wires.
Fig.\ \ref{fig:unstable}(a) shows the real and imaginary parts of
the surface phonon frequency versus wavevector for a typical unstable wire.
One mode $b(q_m,t)$
grows exponentially faster than all others in the harmonic approximation,
and thus may be expected to dominate.  For a single Fourier
component $b(q_m)$, the potential energy $U[b(q_m)]$ of the nanowire
can be evaluated for arbitrarily large $b$
using semiclassical perturbation 
theory. 
$U$ may be expanded semiclassically as\cite{Stafford99,Brack97}
\begin{equation}
U=u{\cal V}+\sigma_s {\cal S} -\gamma_s {\cal C} + \delta U.
\label{eq:potential_exp}
\end{equation}
The volume ${\cal V}$, surface area ${\cal S}$, and integrated mean
curvature ${\cal C}$ of the nanowire 
can be calculated for arbitrary deformations
by simple geometric considerations.
Using semiclassical perturbation theory,\cite{Ullmo96,Creagh96,Stafford01}
the electron-shell correction $\delta U$ can again be expressed in terms of
the classical periodic orbits of a disk billiard, leading to an
expression similar to Eq.\ (\ref{eq:gutzwiller}): 
\begin{widetext}
\begin{equation}
\frac{\delta U[b(q_m)]}{L}  =  \frac{2\varepsilon_F}{\pi}
\sum_{w=1}^\infty \sum_{v=2w}^\infty
\frac{a_{vw}(T) f_{vw}}{v^2L_{vw}}
\left[ \left(1+\frac{b(0)}{R_0}\right)\cos(\theta_{vw}) J_0(\phi_{vw})
- \frac{2b(q_m)}{R_0} \sin(\theta_{vw}) J_1(\phi_{vw})\right],
\label{eq:Uofbofq}
\end{equation}
\end{widetext}
where $\theta_{vw}=k_FL_{vw}(1+b(0)/R_0)-3v\pi/2$,
$\phi_{vw}=2k_F L_{vw} b(q_m)/R_0$, $J_i$ is the $i$-th order Bessel
function, and $b(0)$ is related to $b(q_m)$ by Eq.\ (\ref{eq:b0}).
The result is shown in Fig.\ \ref{fig:unstable}(b).
Although the wire is unstable to breakup under a hypothetical long-wavelength
perturbation,
the energy of the fastest growing mode
reaches a minimum at a finite amplitude, suggesting that the surface 
deformation 
{\em saturates}, and that the wire does not break up, but rather
necks down to the next stable radius.
A similar scenario is predicted
under the diffusive dynamics of Eq.\ (\ref{eq:surface_diff}).
An explicit
nonlinear dynamical simulation \cite{Buerki02} confirms these predictions.

\section{Conclusions}
\label{sec:conclusions}

The stability and surface dynamics of metal nanowires 
were investigated in a continuum approach, including electron-shell effects.
A thermodynamic phase diagram for jellium nanowires was derived, which predicts
that cylindrical wires with certain ``magic'' conductance values are
stable with respect to small perturbations up to remarkably high temperatures.
On the stability boundary, the surface exhibits critical fluctuations, and
the nanowire becomes inhomogeneous.
Both surface phonons and surface self-diffusion of atoms were included in
the linearized surface dynamics. 
It was found that inertial dynamics (phonons) always
dominate the long-wavelength behavior, including the critical points.  
(It must be emphasized, however, that this conclusion holds only for {\em small}
perturbations of the surface.  The irreversible
diffusion of surface atoms is undoubtedly crucial for large-scale surface
deformations.\cite{Buerki02})
A novel reentrant behavior was predicted, in which a straight wire is 
stabilized at intermediate temperatures, but undergoes phase separation into
thick and thin segments as the temperature is lowered.
Finally, for unstable wires, the surface deformation was found to grow
exponentially, dominated by a single Fourier component, and to
saturate at a finite amplitude, indicating that unstable wires may not
break, but rather neck down to the next stable radius.

The results presented in this article
should be directly relevant for nanowires made of monovalent metals,
especially the alkali metals and gold, for which there is clear
experimental evidence of electron-shell effects.\cite{Yanson99,Yanson01,Diaz03}
Moreover, this
simple model may provide qualitative insight into the {\em generic}
surface
properties of metal nanowires, which could guide investigations of more
realistic, material-specific models.

\begin{acknowledgments}
We acknowledge fruitful discussions with J.\ B\"urki and R.\ Goldstein.
CHZ and CAS were supported by NSF grants DMR0072703 and DMR0312028.
This research was supported by an award from Research Corporation.
\end{acknowledgments}

\appendix

\section{The ionic kinetic energy}
\label{sec:appendix}

Here we present some details of the derivation of the expression
(\ref{eq:ke}) for the kinetic energy of the ions.
Inserting Eq.\ (\ref{eq:velocity_potential}) into the first line of
Eq.\ (\ref{eq:ke}) and performing the $z$-integral, one obtains
\begin{equation}
K  = 
\pi\rho_i L
\sum_n\int_{0}^{R_0} \!\!\!r dr \left[q_n^2I_0^{\prime 2}(q_nr)
+q_n^2I_0^2(q_nr)\right] |d(q_n,t)|^2.
\label{eq:step1}
\end{equation}
Using the relation
$I^{\prime}_0(x)=I_1(x)$ 
and the identity
\begin{equation}
 \frac{d}{dx}\left[xI_m(x)I^{\prime}_m(x)\right]=
x\left[I^{\prime2}_m+\left(1+\frac{m^2}{x^2}\right)I^2_m\right],
\end{equation}
the radial integral in Eq.\ (\ref{eq:step1}) may be performed, leading to
the result
\begin{equation}
K =  \pi\rho_iL\sum_{n=-N}^Nq_nR_0I_0(q_nR_0)I_1(q_nR_0)|d(q_n)|^2.
\label{eq:step2}
\end{equation}
Finally, eliminating  $d(q_n,t)$ from Eq.\ (\ref{eq:step2})
using the relation (\ref{eq:dandb}), one obtains the second line of
Eq.\ (\ref{eq:ke}).




\end{document}